\documentclass[sigconf, nonacm]{acmart}

\usepackage[utf8]{inputenc}
\usepackage{graphicx}
\usepackage{xcolor}
\usepackage{balance}
\usepackage[inline]{enumitem}
\usepackage{cancel}
\usepackage{tikz}
\usetikzlibrary{decorations.pathreplacing,calligraphy}
\usepackage{appendix}
\usepackage[compact]{titlesec}

\newlist{todolist}{itemize}{2}
\setlist[todolist]{label=$\square$}
\usepackage{enumitem}
\setitemize{noitemsep,topsep=0pt,parsep=0pt,partopsep=0pt}
\usepackage[capitalise]{cleveref}
\Crefname{equation}{Eq.}{Eqs.}
\Crefname{figure}{Fig.}{Figs.}
\Crefname{section}{Sec.}{Sections}

\renewcommand\footnotetextcopyrightpermission[1]{} 
\begin{document}

\author{Mohammad Reza Saleh Sedghpour}

\affiliation{%
   \institution{Department of Computing Science, Ume{\aa} University}
   \city{Ume{\aa}}
   \country{Sweden}}
\email{msaleh@cs.umu.se}

\author{Alessandro V. Papadopoulos}
\affiliation{%
   \institution{M{\"a}lardalen University}
   \city{V{\"a}ster{\aa}s}
   \country{Sweden}}
\email{alessandro.papadopoulos@mdu.se}

\author{Cristian Klein}
\affiliation{%
   \institution{Department of Computing Science, Ume{\aa} University}
   \city{Ume{\aa}}
   \country{Sweden}}
\email{cklein@cs.umu.se}

\author{Johan Tordsson}
\affiliation{%
   \institution{Department of Computing Science, Ume{\aa} University}
   \city{Ume{\aa}}
   \country{Sweden}}
\email{tordsson@cs.umu.se}

\title{Artifact Evaluation for Distributed Systems: Current~Practices~and~Beyond}



\begin{abstract}
Although repeatability and reproducibility are essential in science, failed attempts to replicate results across diverse fields made some scientists argue for a reproducibility crisis. In response, several high-profile venues within computing established \emph{artifact evaluation tracks}, a systematic procedure for evaluating and badging research artifacts, with an increasing number of artifacts submitted.

This study compiles recent artifact evaluation procedures and guidelines to show how artifact evaluation in distributed systems research lags behind other computing disciplines and/or is less unified and more complex. We further argue that current artifact assessment criteria are uncoordinated and insufficient for the unique challenges of distributed systems research. We examine the current state of the practice for artifacts and their evaluation to provide recommendations to assist artifact authors, reviewers, and track chairs. We summarize the recommendations and best practices as checklists for artifact authors and evaluation committees. Although our recommendations alone will not resolve the repeatability and reproducibility crisis, we want to start a discussion in our community to increase the number of submitted artifacts and their quality over time.

\end{abstract}

\maketitle

\section{Introduction}\label{introduction}
Repeatability and reproducibility are critical components of the scientific process because they prevent disseminating flawed results and allow us to rely on reported findings~\cite{repeatability_in_cs}. Many authors have defined reproducibility and repeatability and related terms in slightly different ways~\cite{abedi2015conducting,drummond2009replicability,feitelson2015repeatability,jasny2011again}. The ability to repeat an experiment using the same procedure on a system that is identical or comparable to that used originally and obtain results identical or very similar to those reported is referred to as \emph{repeatability}. In contrast, \emph{reproducibility} refers to the ability for a scientific hypothesis to be confirmed independently by a separate team~\cite{vivtek2011repeatability}. Although these concepts are clearly related and might even be considered identical in some contexts, the distinction between them is important in computer science. Regardless, there is clear evidence that many research results in different fields of science cannot be reproduced~\cite{reproducibilty, hermann2022has}, and 90\% of researchers who completed an online questionnaire on reproducibility stated that there is a reproducibility crisis~\cite{Baker2016} and only 2.4\% of publications in one of the top software engineering conferences are associated with an artifact and there are no replicated or reproduced results in the last decade~\cite{baldassarre2023re}. Furthermore, other studies have shown that the peer-review process cannot ensure reproducibility on its own ~\cite{repeatability_in_cs,software_crisis,delling_et_al:DR:2016:6146}.

One strategy to make research more reproducible and repeatable is to give researchers extra incentives to publish their findings with evidence of reproducibility~\cite{acm_AE_policy}. Accordingly, several conferences and journals have established a systematic artifact assessment and badging process known as the \emph{Artifact Evaluation Track} (AET) to emphasize the importance of reproducibility in experimental research~\cite{krishnamurthi2013artifact}. Based on our study and our experiences as both artifact evaluation committee members and authors, the adoption of this approach is demonstrated by a growth from a mere 8 artifacts submitted across the entire field of computer science in 2015, to 614 artifacts submitted in 2021.  

The motivations for submitting artifacts for evaluation are diverse. Some researchers see artifacts as \emph{supplementary material} needed to convince reviewers that the reported results were obtained in good faith, at least during the analytical phase of their work, particularly in cases where the infrastructure required for data collection may be unavailable to most other researchers. Others see artifacts as a way of speeding up research as old artifacts can be reused in new experiments, similar to how, once written, a software library can accelerate any number of future software development projects. There is also anecdotal evidence that articles including an evaluated artifact receive more attention from the scientific community~\cite{childers2017artifact,heumuller2020publish}.

When talking about Artifact Evaluation (AE), it is important to distinguish between two approaches to empirical evaluation: simulation and emulation. Simulation involves writing a program that models a computing system and the proposed solution, whereas emulation involves setting up a scaled-down version of a real computing system. Emulation is usually more time-consuming than simulation. The two approaches can complement each other - for example, one could run simulations to sweep a large parameter space and then perform emulation using a subset of parameters to validate the simulation results.

To clarify the current state and challenges of artifacts and procedures for their evaluation in distributed systems research, this study seeks to answer three research questions:
\begin{enumerate}[label=\textbf{RQ\arabic*}.,leftmargin=0.053\textwidth]
    \item \emph{What is the current state of the practice for AE in the distributed systems field?}
    \item \emph{How unified are existing AE practices in distributed systems research?}
    \item \emph{How can reproducibility, artifacts, and AE practices be improved and promoted in distributed system conferences and journals?}
\end{enumerate}

To answer these questions, we gathered data on recently disclosed AE procedures and guidelines (RQ1, see overview in \cref{tab:venues}). Our findings show that AE practices in the field of distributed systems (the research area of our group) either lag behind those used in other fields of computer science (RQ2) or are less unified and more complex (see \cref{tab:criteria}). Indeed, AE practices in distributed systems seem limited and focused mainly on simulation. This is expected because emulating a distributed system is costly in terms of both infrastructure footprint and in terms of human time because setting up emulation testbeds is often difficult to fully automate. 

In this paper, we argue that current AE guidelines are insufficient to capture the unique challenges presented by distributed systems. We start by comprehensively reviewing the history of efforts to improve reproducibility and repeatability in computer science research (see \cref{background}). We then describe the approach used to characterize AE practices in this work, including the study process, results, and threats to the validity of our results (see \cref{approach}). We then review recent AE procedures (but not the evaluation of particular artifacts) and highlight challenges unique to distributed systems (see \cref{best-practices}). We conclude by making recommendations to help artifact authors, artifact reviewers, and AET chairs (see \cref{our-recommendations}). While many seem obvious or trivial, we bring evidence that these recommendations are not universally followed (see \cref{tab:criteria}).
Furthermore, we believe these recommendations are specific to distributed systems research, as they may lead to a sound and repeatable way for artifact implementation and evaluation. In contrast, other studies highlighted such recommendations primarily for experiment designing and result reporting~\cite{papadopoulos2019methodological}. Finally, we summarize our main findings and recommendations (see \cref{discussion}) as checklists for both artifact authors and AE committees (see \cref{appendix-a,appendix-b}). 

The main \textbf{contributions} of this study are:
\begin{enumerate}
    \item We gather the existing state of the practice and AE procedures reported at distributed systems conferences (RQ1, see \cref{best-practices}).
    \item Our analysis shows that the existing state of the practice is not applied uniformly at all conferences (RQ2, see  \cref{tab:venues}). 
    \item We offer recommendations to mitigate challenges and complexities and promote repeatability and reproducibility in distributed systems research (RQ3, see \cref{our-recommendations}).
    \item We provide checklists for both artifact authors and AE committees (\cref{appendix-a,appendix-b}).
\end{enumerate}

While we do not claim to have found the silver bullet, we hope that this paper will serve as a conversation starter in the distributed systems community and incrementally lead to more and better artifacts and a fairer AE process. We believe that our article will be useful to jump-start the work of Ph.D. students who want their distributed systems research to be reproducible.


\section{Background}\label{background}

This section discusses previous studies relevant to AE and reproducibility. The concepts of reproducibility and repeatability have been applied to research in all fields of science in which experiments play a role, including biology~\cite{vaux2012replicates}, psychology~\cite{open2015estimating}, and scientific computing~\cite{easterbrook2014open,fomel2008guest}.
 Various studies have therefore been performed to find ways of improving reproducibility and repeatability, leading  
to the introduction of best practices for researchers such as making code available, providing basic and/or extensive documentation, automation and testing, accessibility, modularity, software management, and licensing~\cite{fehr2016best}. In addition, some publishers and standard organizations have tried to push boundaries by giving researchers extra incentives to make research publications reproducible~\cite{acm_AE_policy,niso2021}. 

These efforts resulted in the development of an intermediate Artifact Evaluation process by the Association for Computing Machinery (ACM). This process was introduced in 2011 and is now widely adopted by conference organizers~\cite{childers2016artifact,krishnamurthi2013artifact}. It requires all research artifacts to pass a rigorous audit~\cite{acm_AE_policy}. In this context, an \emph{artifact} is defined as a digital item developed by the authors of a publication that was either used in the authors' study or generated during their experiments. Three different artifact review badges can be awarded to research articles published in ACM publications: Artifacts Evaluated, Artifacts Available, and Results Validated. These badges are considered distinct: any one, two, or all three can be assigned to any particular work, depending on the review procedures established by the journal or conference. 

    \emph{Artifact Evaluated}: This badge is given to papers whose accompanying artifacts have passed an independent audit. This badge does not require that artifacts be made publicly available. They must, however, be made available to reviewers. It has two separate grades, only one of which may be awarded to a given publication:
    \begin{itemize}[leftmargin=*]
        \item \emph{Functional}: The artifacts are determined to be documented, consistent, complete, usable, and to provide acceptable evidence of verification and validation.
        \item \emph{Reusable}: The artifacts linked with the article substantially exceed the criterion of minimum functioning. That is to say, they possess all of the characteristics required for the above Functional grade but are also carefully documented and well-structured to a level that enables reuse and repurposing.
    \end{itemize}
    
    \emph{Artifact Available}: This badge is given to papers with linked artifacts that are permanently available for retrieval.
    
    \emph{Results Validated}: This badge is awarded to papers whose main scientific results have successfully been obtained by an individual or team other than the author. It has two grades:
    \begin{itemize}[leftmargin=*]
        \item \emph{Results Reproduced}: The paper's main findings were independently achieved in subsequent work by an individual or team other than the authors, partly using artifacts supplied by the authors.
        \item \emph{Results Replicated}: The paper's main findings were independently achieved in subsequent work by an individual or team other than the authors without using author-supplied artifacts.
    \end{itemize}
Moreover, recently, IEEE Computer Society introduced a reproducibility badging system that offers three different badges: Code/Data Available, Code/Data Reviewed, and Results Reproducible\cite{reproducibility_IEEE_CS,tpds_reproducibility}.
    \emph{Code/Data Available}: This badge indicates that the code, including any associated data and documentation, provided by the authors is reasonable and complete and can potentially be used to support the reproducibility of the published results.
    
    \emph{Code/Data Reviewed}: This badge demonstrates that the code, including any associated data and documentation, provided by the authors is reasonable and complete, runs to produce the outputs described, and can support reproducibility of the published results.
    
    \emph{Results Reproducible}: This badge was recently introduced, and we could not find any official definition of it.

An \emph{Artifact Evaluation Track} (AET) based on these definitions and badges has been applied at many conferences to foster reproducibility. This was considered necessary because multiple studies have highlighted the reproducibility crisis, showing that many authors cannot even replicate their own results after a year~\cite{dietrich2015dataref}, let alone enable others to reproduce them many years in the future. For instance, Collberg and Proebsting attempted to reproduce the results of several papers published at different conferences~\cite{collberg2016repeatability} but found that they could only reproduce 32.3\% of the published results without communicating with authors, and 48.3\% percent after such communication. Papers written by authors with industry ties had lower reproducibility rates. They also found that some researchers were unwilling to share code and data, and those who did share often provided insufficient information to reproduce the experiment. Collberg also investigated the repeatability of 402 ACM papers published at computer systems conferences~\cite{collberg2015repeatability}, taking the ability to obtain and build the source code in a reasonable amount of time as a minimum acceptable level of reproducibility. The author's assistants could only do this within 30 minutes in 32.3\% of the studied cases; 15.9\% of the artifacts required more than 30 minutes, and 5.7\% required additional but still reasonable efforts. However, the code failed to build in 2.2\% of the cases, authors refused to supply code in 7.5\% of the cases, and 36.3\% of authors simply never responded to code-sharing requests. 

Because of these problems, several groups have sought to mitigate the repeatability and reproducibility crisis in computer science. Herrmann et al.~\cite{hermann2020community} assessed the efficacy of AEs in software engineering and programming language conferences and identified possible improvements for these processes. Berger et al.~\cite{berger2019checklist} observed that papers with significant methodological flaws were published at leading programming language conferences (SIGPLAN) and, therefore, formed an ad hoc committee to develop an empirical evaluation checklist for authors and reviewers of research on programming languages. Similarly, the National Academies of Sciences, Engineering, Medicine (NASEM) published a report that defined reproducibility and replicability and examined factors that may lead to non-reproducibility and non-replicability in research~\cite{national2019reproducibility}. This report also provides recommendations for researchers, academic institutions, journals, and funding agencies concerning steps they can take to improve reproducibility and replicability in science. Barowy et al.~\cite{barowy2022howtoAEC} proposed artifact guidelines based on their experiences as both AE committee members and authors. Papadopoulos et al.~\cite{papadopoulos2019methodological} proposed an experimental methodology that focuses on selecting suitable metrics and the reproducibility of experimental results. They presented eight principles for increasing the reproducibility of performance experiments on cloud computing platforms. Finally, cTuning is a highly active foundation that aims to develop a methodology and open-source framework for collaborative and reproducible Machine Learning systems research~\cite{fursin2020enabling,fursin2021collective}. This foundation also provides artifact submission and evaluation guidelines using a so-called Artifact checklist~\cite{fursin2022ctuning}.

In another study, the authors assessed the current state of artifact availability and reproducibility in leading computer networking conferences. They surveyed papers published in 2017 to explore the importance of reproducibility and encourage researchers to prioritize ease of reproduction~\cite{Survey2018flittner}. Moreover, Malik examines the benefits and challenges of the artifact description and artifact evaluation (AD/AE) process in systems research conferences. She highlights the importance of increased transparency, data management, and reproducible containers in facilitating the validation and reproducibility of research claims~\cite{Artifact2021Malik}. Krishnamurthi et al. report their experience as chairs of the AE process at a software engineering conference in 2011, mainly discussing their personal opinions about their experiences in that role~\cite{krishnamurthi2013artifact}.

Bajpai et al.~\cite{bajpai2019dagstuhl} provide a guideline focusing on the importance of experiment design in enhancing reproducibility and establishing a solid groundwork for future research in experimental disciplines such as Internet measurements and networked systems. Additionally, another paper on computer networking discusses the importance of reproducibility of research results. It highlights the lack of conviction in the reproducibility of a significant portion of research findings and the efforts made by the ACM SIGCOMM community to address this issue through new policies and workshops~\cite{Saucez2018thoughts}.

\medskip
Despite the recent interest in promoting reproducibility and repeatability, few studies have investigated reproducibility in computer systems research~\cite{Timperley2021artifact-sharing,heumuller2020publish,chen2022towards,papadopoulos2019methodological}. This study differs from previous works on reproducibility in computer systems research in that it focuses on the challenges presented by distributed systems and intends to serve as a conversation starter in the distributed systems community to highlight the importance of reproducibility and repeatability. 

\section{Challenges of distributed systems research}\label{challenges}


Distributed systems research aims to study a collection of computing elements that appear to its users as a single coherent system. Those computing elements could be either a hardware device or a software process that collaborate to be seen as a single system~\cite{ds-definition}. However, the unique characteristics of distributed systems research can present challenges when reproducing the artifacts and results of such research work, and it may not be easy to fully automate the experiments and artifacts. In this section, we discuss these challenges in detail.

\begin{enumerate}[label=\textbf{C\arabic*}.,wide, labelwidth=!, labelindent=0pt]
    \item \emph{Hardware includes everything:\label{c1}} One significant challenge in distributed systems research lies in the hardware requirements, which can vary widely. Researchers often need access to diverse hardware resources for computation, storage, and networking. Although some cloud platforms, supercomputers, and simulators offer various hardware configurations, it can be challenging for reviewers and readers to replicate the original research environment precisely. \\ This challenge arises from the need for exact hardware matching. For example, specific features like hardware counters to measure power consumption or the ability to control the working frequency of resources might not be uniformly available across all platforms or simulators. The diverse nature of hardware requirements makes it difficult for reviewers and readers to accurately reproduce the artifact, as they may need to seek out or emulate very specific hardware configurations.

    \item \emph{Software stack:\label{c2}}
       Another challenge in distributed systems research revolves around the employed software stack. Researchers often explore aspects of a distributed software stack that contains a wide range of components, from hypervisor control parameters to application code. These components can have considerable influence on the results of experiments. However, the research paper alone may not sufficiently communicate the obstacles that reviewers and readers might face when attempting to replicate the research work. The main concern lies in the diverse software and run-time environments commonly used in distributed systems research. These environments can incorporate programmable application-aware networking tools, various operating systems, and libraries for experiment execution. \\Replicating the precise software stack utilized in the original research can be challenging. Achieving the necessary precision level often requires matching specific versions, configurations, and dependencies. Consequently, the complicated nature of the software stack can make it difficult for reviewers and readers to reproduce the artifact accurately.


    \item \emph{Large scale environment:\label{c3}} Reproducibility in distributed systems research is further complicated by the requirement for a large-scale environment. This complexity is caused by the need for experimenters to access expensive and diverse resources to conduct various experiments. These large-scale environments contain a wide range of components, including bare-metal servers or virtual machines with varying numbers of CPUs or GPUs, storage devices of various sizes, and a diverse networking infrastructure that includes routers and switches.\\ The need for large-scale environments sets distributed systems research apart from many other fields within computer science. Unlike some domains where experiments can be conducted on individual machines or smaller clusters, distributed systems research necessitates a more extensive environment. This scale introduces unique challenges in terms of resource allocation, coordination, and management.\\Thus, reproducing the large-scale environment used in the original research presents significant challenges for reviewers and readers. Acquiring and configuring an equivalent setup, which includes the precise combination of resources and networking devices, can be demanding and resource-intensive. Moreover, the availability and access to such a large-scale environment may be limited, making it even more challenging to reproduce the artifact accurately.


    \item \emph{Heterogeneity:\label{c4}} Some research endeavors specifically explore the impact of heterogeneity to touch real-world scenarios and require diverse resources to conduct their experiments. Heterogeneity in distributed systems includes variations in hardware, software, configurations, and other parameters across different system components. This introduces complexities beyond those in homogeneous environments. Thus, reviewers and readers attempting to reproduce such research work may need to address the complexities of coordinating and managing diverse resources, each with its capabilities and characteristics.
    
    
    \item \emph{Self-tuning mechanisms:\label{c5}} Reproducibility in distributed systems research is further complicated by autonomous self-tuning mechanisms and control loops that operate independently of the experimenter. These mechanisms, such as CPU throttling for thermal management or background storage re-balancing, are designed e.g., to prevent faults and optimize performance. However, in a distributed environment, cascaded self-tuning mechanisms can introduce additional complexities to the experimental setup. Moreover, these self-tuning mechanisms can interact with one another, creating complicated control loop interactions that can impact the behavior and outcome of experiments. These mechanisms often operate beyond the direct control of the experimenter, making it challenging to precisely reproduce the conditions in which the original artifact was evaluated.
    

    \item\emph{Demarcation:\label{c6}} In distributed systems research, experimenters have to balance their control over the experimental setup with the constraints of the distributed environment, which introduces complexities in defining precise boundaries. This demarcation challenge involves aspects like system component behavior, node interactions, and external factors. Researchers need to consider inherent limitations, uncertainties, and potential dependencies on external services or infrastructure. Determining these boundaries can be challenging for reviewers and readers and may impact their reproduction process.
    

    \item \emph{Nondeterministic behavior:\label{c7}} In distributed systems, the ordering of events can vary due to factors such as network latency, message propagation delays, or asynchronous processing. These inherent characteristics of distributed systems can pose challenges for reviewers and readers attempting to reproduce the research findings despite the apparent simplicity of the example provided. Even when experiment parameters remain constant, the nondeterministic nature of event ordering can lead to subtle differences in the obtained results between different experiment executions. For example, consider an experiment with a load generator and an application with a database backend. Assume that the workload generator generates the same requests at the same times, with millisecond accuracy. However, the system's distributed nature means that a database service located downstream in the call chain will likely receive these queries in a different order for each experiment.
   
\end{enumerate}
The combined effects of these challenges, particularly in large-scale environments, have caused distributed systems research to lag behind other disciplines within the field of computer science, as other disciplines are normally executed on a smaller scale. While various computer science domains, such as the ones that are using or studying storage, may also require specific hardware resources like Persistent Memory, remote direct memory access (RDMA), and others, the unique complexity of distributed systems described above poses additional hurdles for repeatability and reproducibility. In distributed systems research, we often encounter long-term repeatability challenges, particularly due to the utilization of hardware and software resources that depreciate over time. Notably, we do not consider the latter a challenge for AE, as it tends to focus on the recently conducted research.


\section{Surveying state of the practice}\label{approach}
In this section, we introduce our approach to data collection and analysis. We then discuss factors that could reduce the study's validity and the measures taken to ensure validity. 


\begin{table}
\vspace*{-\baselineskip}
    \centering
    \caption{List of keywords used to find conferences and journals related to distributed systems.}
    \resizebox{0.5\columnwidth}{!}{
    \begin{tabular}{|c|l|}
    \hline
        \multicolumn{1}{|c|}{\textbf{No}} & \multicolumn{1}{c|}{\textbf{Keyword}}\\ \hline
        1  & Distributed systems       \\ \hline
        2  & Cloud computing           \\ \hline
        3  & Edge computing            \\ \hline
        4  & Fog computing             \\ \hline
        5  & Serverless computing               \\ \hline
        6  & Service-oriented architectures \\ \hline
        7  & Microservices             \\ \hline
    \end{tabular}
    }
    \label{tab:keywords}
\vspace*{-\baselineskip}
\end{table}

\subsection{Study Subjects}
Our study aimed to review AE procedures used at distributed system conferences and in the field's journals. We chose this method rather than searching for the field's papers due to time limits, and we aimed to extract the existing best practices employed at the field's conferences and journals. To this end, we first needed to identify relevant conferences and journals with Artifact Evaluation Tracks (AETs). We, therefore, searched four different sources to review AE procedures: 
\begin{enumerate*}
    \item ACM Digital Library~\cite{acm_dl},
    \item Xplore tool~\cite{ieee_xplore} by the Institute of Electrical and Electronics Engineers (IEEE),
    \item WikiCFP~\cite{wikicfp},
    \item Researchr~\cite{researchr}.
\end{enumerate*} 

ACM Digital Library was used to search for available artifacts, while the IEEE Xplore tool allowed us to query for supplemental materials such as code or data sets. WikiCFP is a semantic wiki for Calls For Papers in science and technology, while Researchr is a platform for finding, collecting, sharing, and reviewing scientific publications.

We chose to review only conferences and journals that have websites containing a call for artifacts and an AET and have at least one of the keywords listed in \cref{tab:keywords} in their call for papers.

\subsection{Study Process}

To get the names of conferences and journals with an AET, we performed searches for \emph{artifact evaluation track} in WikiCFP and Researchr. We queried ACM Digital Library and IEEE Xplore and filtered all papers containing artifacts and supplemental materials, resulting in the identification of 3154 papers, including such materials that had been published as of August 2022. We then extracted and listed the names of the conferences and journals in which these papers had been published. 
We found 108 conferences or journals with procedures for AE or inclusion of supplemental materials. \cref{fig:current_AE} shows the distribution of the retrieved articles over data sources (journals and conferences) with AETs and years of publication (2016-2022). A simple inspection of this figure shows that AE processes started attracting appreciable attention in 2016. While some artifacts were made available before 2016, they were generally not evaluated by any committee.

Finally, we extracted the conferences whose scopes correspond to the keywords listed in \cref{tab:keywords} and reviewed the procedures of those held in the last three years (see \cref{tab:venues}). We chose to focus only on conferences held in the last three years to reduce the time needed to review the AE procedures and because it was expected that the procedures applied at these more recent conferences would incorporate the lessons learned during earlier conferences. However, we note that the decision to examine only more recent conferences can potentially reduce the validity of our findings (see \cref{threats}).

\begin{table*}
    \centering
    \caption{List of distributed systems conferences with an AE process.}
    \resizebox{\textwidth}{!}{
    \begin{tabular}{|l|l|}
    \hline
        \multicolumn{1}{|c|}{\textbf{Conference Name}} & \multicolumn{1}{c|}{\textbf{Years}}\\ \hline
        International Conference on Architectural Support for Programming Languages and Operating Systems (ASPLOS) &
        2022, 2021, 2020       \\ \hline
        International Conference on Computing Frontiers (CF) & 
        2020                   \\ \hline
        International Conference on emerging Networking EXperiments and Technologies (CoNEXT) & 
        2022, 2021, 2020       \\ \hline
        Joint European Software Engineering Conference and Symposium on the Foundations of Software Engineering (ESEC / FSE) & 
        2021, 2020             \\ \hline
        EuroSys Conference & 
        2022, 2021             \\ \hline
        International Conference on Performance Engineering (ICPE) &
        2022, 2021, 2020       \\ \hline
        International Conference on Software Engineering (ICSE) &
        2022, 2021, 2020       \\ \hline
        International Symposium on Microarchitecture (MICRO) &
        2021                   \\ \hline
        International Middleware Conference &
        2022, 2021, 2020       \\ \hline
        International Conference for High-Performance Computing, Networking, Storage, and Analysis (SC) &
        2022, 2021, 2020       \\ \hline
        Symposium on Operating Systems Principles (SOSP) &
        2021                   \\ \hline
        Annual Symposium on Principles and Practice of Parallel Programming (PPoPP) &
        2022, 2021, 2020       \\ \hline
        International Conference on Utility and Cloud Computing (UCC) &
        2021                   \\ \hline

    \end{tabular}
    }
    \label{tab:venues}
\vspace*{-\baselineskip}
\end{table*}

\begin{figure}[!tb]
\vspace*{-\baselineskip}
    \centering
    \includegraphics[width=\linewidth]{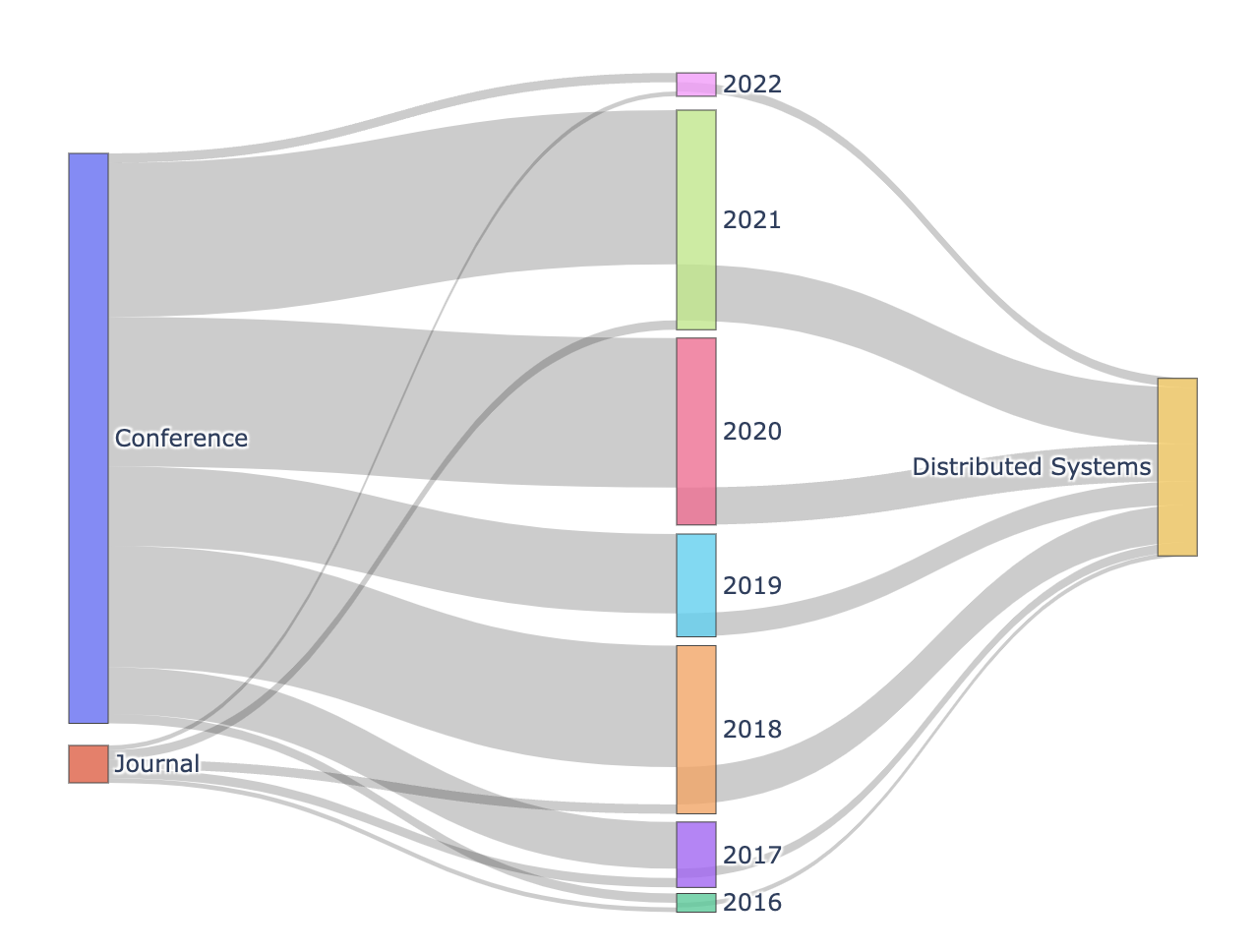}
    \caption{Conferences and journals with AETs.}
    \label{fig:current_AE}
\vspace*{-\baselineskip}
\end{figure}

\subsection{Analysis}

Our initial goal in the analysis was to identify the state of the practice used in recent AE procedures for distributed systems research. To this end, we reviewed all of the practices for artifacts and procedures for their evaluation described on the public websites of the targeted conferences, which are listed in \cref{tab:venues}. 

For each conference and year, we cross-referenced and categorized the current practices and lessons learned documented in the corresponding calls for artifacts. All recommendations provided by the studied conferences are listed in \cref{tab:criteria}. 

\begin{table*}
\centering
\caption{Summary of recommendations for practices for artifacts and procedures for their evaluation at selected conferences. Check marks indicate conferences that provided recommendations. Dashes indicate conferences where recommendations were either not provided or not available online.}
\begin{tabular}{l}
    \rotatebox{0}{\includegraphics[width=\textwidth]{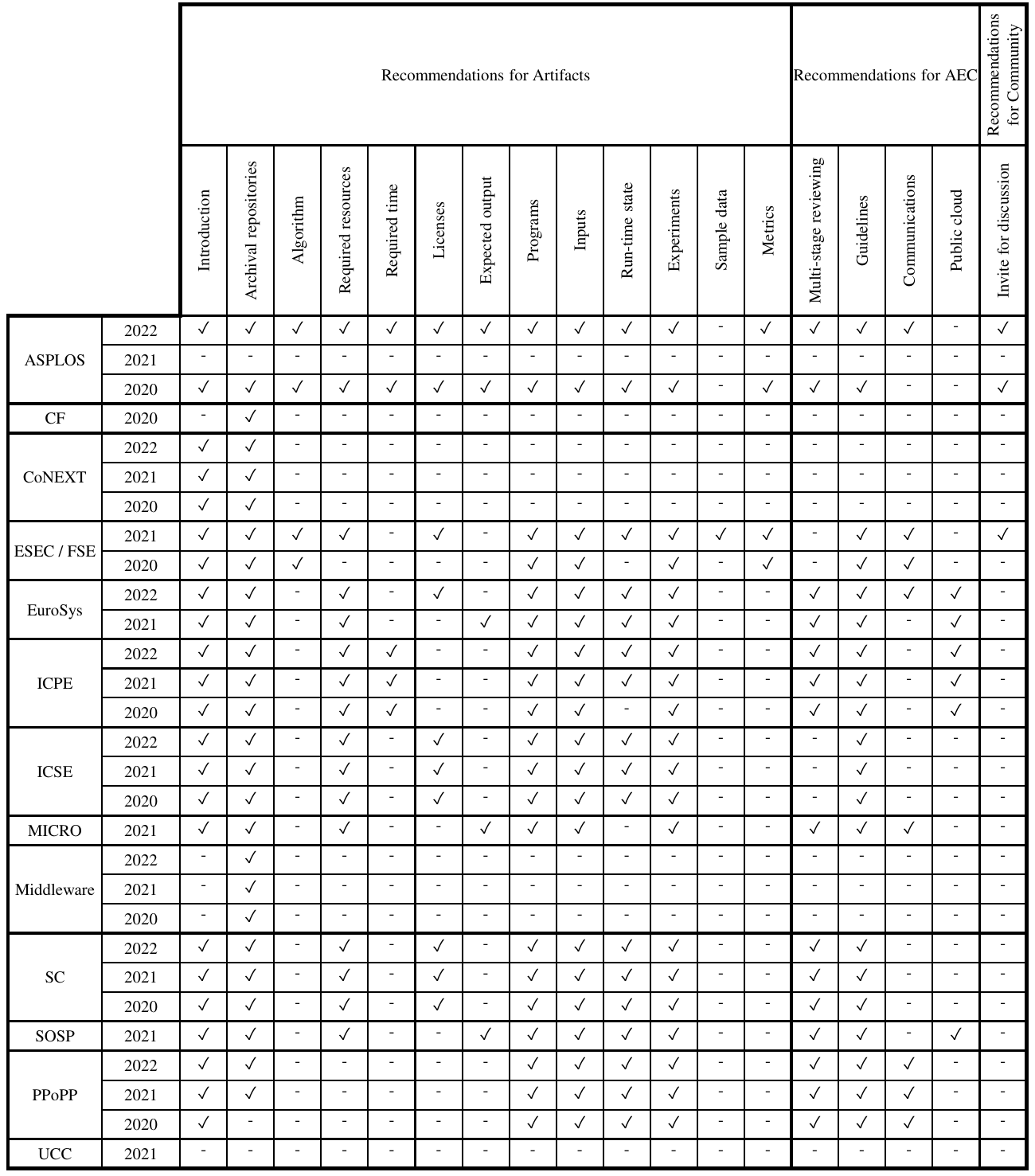}}
\end{tabular}
\label{tab:criteria}
\vspace*{-\baselineskip}
\end{table*}

\subsection{Threats to Validity}\label{threats}
Despite careful research design, studies such as those presented here inevitably have limitations that may reduce their validity. Such limitations for this work are summarized below.

\subsubsection{External Validity}
Only conferences held within the three years preceding August 2022 were examined in this work. This can be justified using an evolutionary argument: any practices adopted in earlier conferences but subsequently discontinued are unlikely to have been very effective --- if they had been effective, they would not have been discontinued. In essence, the last three years should capture the state of the practice.  In addition, thoroughly reviewing earlier conferences would have been very time-consuming. However, the decision to focus only on the last three years means that the findings presented herein cannot be directly generalized to earlier conferences, particularly as guidelines may differ between years, even in cases where the AE committee was unchanged. Another potential limitation is that we only searched four repositories to identify relevant AETs. While these repositories are likely to include all relevant conferences and journals, this decision means that our conclusions cannot be applied to all conferences and journals and may be biased due to sampling a small subset of a larger population. To summarize, while we have no direct evidence that our results represent the current state of the practice and procedures of all conferences and journals, we expect that similar results would be obtained if different conferences and journals were chosen.

\subsubsection{Internal Validity}
Internal validity is inevitably affected because some decisions must be made when choosing search keywords. An empirical analysis does not suggest that the results obtained would have been radically different if alternative keywords had been used, but this is impossible to prove. Our study is based only on AE processes with public written descriptions. However, additional information could have been passed during some AE processes that would not have been written down or would not be publicly available. We deem this unlikely given the long tradition of written instructions for paper review and the concerns about reproducibility and replicability that inspired the concept of AE. Nevertheless, this can be seen as a limitation of our study and a reminder of the importance of publicly available evaluation procedures.

\section{Existing state of the practice}\label{best-practices}
To improve the state of repeatability and reproducibility in distributed systems research, we review the practices for artifacts and procedures for their evaluation that are applied at the recent conferences listed in \cref{approach}. The first general insight from this review is that around half of the studied conferences followed the recommendations of the cTuning foundation~\cite{ctuning}. While the recommendations outlined in the studied conferences are comprehensive, a key issue, as discussed in \cref{our-recommendations}, lies in their effective execution. We propose that even if these existing practices are followed, challenges or insufficiencies may persist when attempting to reproduce the artifacts, as evidenced by experience from past conferences and journals. The recommendations provided at the studied conferences are summarized in \cref{tab:criteria}.

\subsection{Recommendations for Artifact Authors}
In the same way that a research paper should have a well-defined structure and progression from the abstract to the conclusion, an ideal artifact should be divided into different sections to facilitate the reproduction and repetition of the research. 

\subsubsection{Introduction}
The introduction section comprises the first portion of the artifact. It should describe the artifact and how it supports the research paper. This section could have other names, such as abstract. It may also be helpful to include a brief description of the minimum hardware and software requirements, the validation procedure, and the expected results to guide the selection of appropriate reviewers for the artifact.

\subsubsection{Archival Repositories}
Sharing data such as artifacts makes research more transparent and allows others to assess and re-evaluate results more easily and cheaply. To facilitate such sharing, Wilkinson et al.~\cite{wilkinson2019fair} proposed the FAIR principles for scientific data management, which state that scientific data should be Findable, Accessible, Interoperable, and Reusable. The FAIR principles have been implemented by several publicly accessible archival repositories such as Zenodo~\cite{zenodo}, FigShare~\cite{figshare}, or Dryad~\cite{dryad}. In addition, the ACM issues its \emph{Artifact Available} badge only to papers with artifacts deposited in a publicly accessible archival repository. When an artifact is deposited in such a repository, it is assigned a Digital Object Identifier (DOI), which should be included in the artifact's description.  

\subsubsection{Algorithm}
Most research papers propose a new algorithm to control a system. In such cases, it is commonly recommended that the proposed algorithm should be described in the artifact's introduction section to help reviewers understand the essence and goal of the implemented artifact.

\subsubsection{Required Resources}
Some experiments might need specific hardware such as a commercial or academic cloud platform, supercomputer, architecture simulator, CPU, GPU, or a neural network accelerator with specific features like hardware counters to measure power consumption and SUDO access to CPU/GPU frequency. Experiments may also need specific software and run-time environments such as specific container orchestration systems, programmable application-aware networking tools, a specific operating system version, or supporting software such as libraries. A specific compiler or transformation tool may also be needed in some cases. Specifying such hardware and software requirements, including their versions, can greatly reduce the incidence of problems during the evaluation and repetition of experiments.We believe that this may address challenges \textbf{C1}, \textbf{C2}, \textbf{C3}, and \textbf{C7}.

\subsubsection{Required Time}
Some research papers discuss sophisticated scenarios that span long periods and involve measurement or analysis on different timescales that may be as long as days, weeks, or even months. Consequently, many recommendations state that the time needed to perform the experiments associated with an artifact should be specified to enable the AE committee to find an appropriate way of evaluating the artifact.

\subsubsection{Licenses}
If the authors intend to make the artifact publicly available, it is recommended that they provide information about any relevant licenses, as the community benefits from understanding how the artifact can be reused. 

\subsubsection{Artifacts' Expected Output and Results}
It has been suggested that the intended outcomes and output format be described to help reviewers prepare and examine the artifact more accurately.

\subsubsection{Programs}
As some artifact authors use specific benchmarks or preconfigured versioned packages, it should be stated in the artifact description whether the benchmark or preconfigured versioned packages are included in the submitted artifact or must be downloaded separately. In the latter case, the version to be downloaded should be clearly specified. In addition, benchmarks or packages may be public or private. If a private benchmark or package not available to reviewers is used, it is recommended that an alternative that reviewers can use in AE should be specified. We believe this may address challenges \textbf{C2} and \textbf{C7}.

\subsubsection{Inputs}
A study may use diverse inputs, including data sets, models, and/or configuration files. Therefore, these inputs should be attached to the artifact to enable the repetition and reproduction of the study and its results. Alternatively, if they must be obtained from external sources, there should be a clear description of how to download and install them. We believe this may address challenges \textbf{C2} and \textbf{C7}.

\subsubsection{Run-time State}
At run-time, an object instance in the used software may have specific attribute values or exist in a particular state. This might cause the artifact to be sensitive to run-time phenomena such as cold or hot cache or contention in the cache or network. Such sensitivities should be analyzed and reported in the artifact's description so the evaluation committee can review it more accurately. We believe this may address challenges \textbf{C2} and \textbf{C7}.

\subsubsection{Experiments}
The artifacts should include clear descriptions of how to prepare and perform the experiments and replicate or reproduce the results. The procedures for reproducing or replicating the results could include shell scripts, details of manual steps to be taken by the user, or automated workflows such as using virtualization techniques like virtual machines or containerization to reduce the experiments costs and time. 

\subsubsection{Sample Data}
Research papers may have two phases: a data-gathering phase during experiments and the subsequent analysis of the gathered data. Gathering data may take a long time or require specific hardware resources that are unavailable to the AE committee. It is therefore recommended that results from data-gathering should be included in the artifact so that the AE committee can at least repeat the analysis phase. The process for retrieving such data and repeating the analysis should be explicitly explained in such cases. We believe this may address challenges \textbf{C2} and \textbf{C7}.

\subsubsection{Metrics}
Metrics are quantitative tools used to help assess and evaluate an experiment's results. A study may use multiple metrics at different stages - for instance, one set of metrics may be used to represent raw measurements of resource usage or behavior. These might be low-level usage summaries provided by the operating system or higher-level data tied to a component's specific functionality or work, such as measurements of throughput or response times by web servers. Some metrics are presented in relation to a total capacity, while others are represented as rates that indicate a component's activity. It is important to describe all metrics to give reviewers a thorough understanding of the technical component under examination.

\subsection{Recommendations for Artifact Evaluation Committees}
The strategy used to manage AE and evaluation processes is another factor that affects the reproducibility and repeatability of experiments and research outcomes in distributed systems research. Recent conference AE committees have used various strategies for this purpose, described below.

\subsubsection{Multi-stage Reviewing}
The majority of conferences used a two-stage reviewing method. The first step, often called Kick-the-Tires, represents the reviewers' initial engagement with the artifacts to determine whether anything needed, e.g., to install and comprehend the artifact, is missing. If something is lacking, the reviewers should request it from the authors during the first round. The second step involves thoroughly evaluating the artifact; the ultimate decision about awarding a badge depends on the results of this stage. As a complex artifact can have minor and significant issues, a multi-stage reviewing process seems to be appropriate for AE in distributed systems research.

\subsubsection{Providing Guidelines}
Many calls for Artifact Evaluation Committee members indicate that participation in such committees is especially suitable for early career researchers such as Ph.D. students, including first-year Ph.D. students. A consequence of this is that some committee members may be unfamiliar with specific distributed systems algorithms or technologies, making the experiments significantly more challenging to execute than they were for the authors. On the other hand, the artifact authors may have a restricted perspective or (as discussed in \cref{background}) may lack the time to provide an artifact accessible to reviewers with different levels of knowledge. Alternatively, when creating their artifacts, the authors may have chosen to use trending technologies that are unfamiliar to committee members to make themselves more attractive to prospective industrial employers~\cite{fritzsch2021resume}, as the study highlights a phenomenon describing the overemphasis of trending technologies in both job offerings and resumes, which may influence hiring decisions in various sectors, including research and education. Therefore, providing detailed and practical guidelines describing an artifact's critical components and ``dos and don'ts'' may help authors submit repeatable and reproducible artifacts. Furthermore, providing detailed and practical guidelines that precisely describe the evaluation criteria may facilitate more accurate AEs. We believe this may address all challenges described in \cref{challenges}.

\subsubsection{Communications}
Subjecting each artifact to several rounds of review makes it vital to maintain anonymous communication between reviewers and authors. However, some platforms that were designed for the evaluation of papers (text) rather than artifacts (computer software) and seem to be incompatible with multi-stage assessments. Furthermore, an inexperienced reviewer may require guidance through the evaluation process. As a result, it is advisable to use a platform that enables anonymous contact between authors and reviewers at all times while also allowing for communication between reviewers and chairs. In addition, communication should be promoted between reviewers and authors. 

\subsubsection{Public and Academic Clouds}
Some artifacts, particularly in the distributed system research domain, may need significant computing or expensive resources, such as special-purpose hardware unavailable to all reviewers. Therefore, offering public cloud credits or access to academic clouds may be useful in evaluating artifacts. We believe this may address challenges \textbf{C1} and \textbf{C7}.

\subsection{Recommendations for the Community}
In light of the challenges inherent in ensuring repeatability and reproducibility in scientific research, particularly within the domain of distributed systems, fostering open dialogue and collaboration within the research community is paramount, and the community should always be open to fresh ideas for improvement in these areas. Conference chairs and community leaders play a vital role in promoting these principles by encouraging active engagement and debate on evaluation approaches. We have seen many conferences where the chairs asked everyone to debate the evaluation approach in an open AE group or on other social media platforms.
\section{Proposed recommendations}\label{our-recommendations}
We believe that implementing all of the recommendations described in the preceding section would greatly improve the AE process and the repeatability and reproducibility of distributed systems research. However, as shown in \cref{tab:criteria}, not all of these recommendations are currently implemented in the conferences. Therefore, existing practices for artifacts and procedures for their evaluation need refinement. In this section, we offer a set of recommendations based on our own experiences of AE procedures that could be added to the existing recommendations discussed in \cref{best-practices}. Some of these recommendations may have been previously presented or used by other research communities, including in distributed systems, but to date, they have not been summarized in the following manner for the benefit of our community.

\subsection{Recommendations for Artifact Authors}
\subsubsection{Plan Artifact Submission in Advance, Not After Paper Acceptance}
Based on our experience, the likelihood of not thoroughly reading the guidelines when submitting an artifact is high, especially as authors are usually notified about the artifact track only after being notified of their paper's acceptance. Consequently, they may only have around 10 days to prepare artifacts and may rush to submit the artifact while simultaneously addressing comments on their original paper. Therefore, we strongly recommend that artifacts for submission be prepared well in advance based on previous conferences' AET guidelines. We believe this may address challenges described in \cref{challenges}.

\subsubsection{Version Control}
Version control systems were originally developed by teams collaborating on large industrial software engineering projects but have also become vital tools for individual publications. However, we have found that some artifacts do not use a version control system. Some of the artifact's features may be inadvertently lost during the AE process. Consistent use of version control systems is, therefore, recommended~\cite{stodden2013best}. We believe that this may address challenges \textbf{C2}, \textbf{C3}, and \textbf{C4}.

\subsubsection{Test Driven Development to Avoid Introducing Bugs During Artifact Evaluation}
Communication between authors and reviewers during the AE process may be necessary to address critical bugs during the initial rounds of evaluation. In such cases, the author's attempts to fix the bugs may cause other parts of the artifact to break because test-driven development procedures are rarely used during artifact creation. To avoid such problems, it is strongly recommended that artifacts should be implemented with procedures to verify their functionality~\cite{tdd-reproducibility}. We believe that this may address challenges \textbf{C2}, \textbf{C3}, \textbf{C4}, \textbf{C5}, \textbf{C6}, and \textbf{C7}.
\begin{figure*}[!ht]
    \centering
    \includegraphics[trim={0 0.9cm 0 3.5cm}, width=0.9\linewidth]{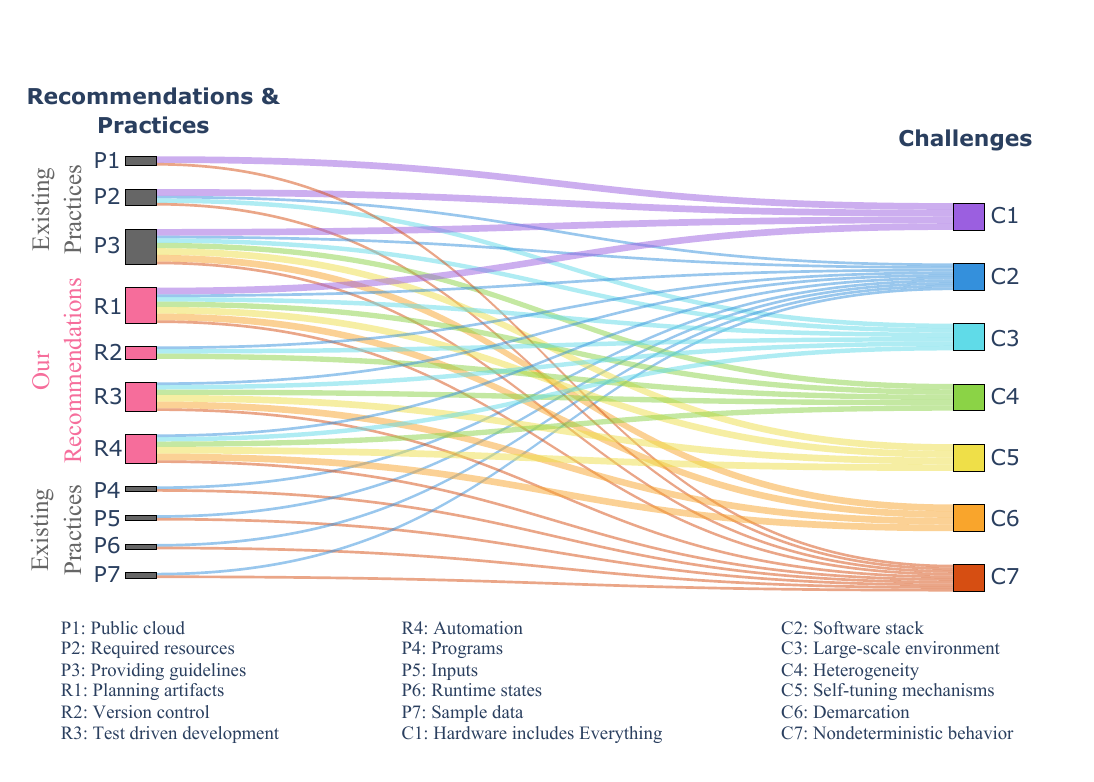}
    \caption{Mapping between challenges (see \cref{challenges}) and the existing state of practices (shown in gray, see \cref{best-practices}) and our recommendations (shown in pink, see \cref{our-recommendations}). Some of the practices and recommendations do not impact the challenges. }
    \label{fig:chal_rec_mapping}
\vspace*{-\baselineskip}
\end{figure*}

\subsubsection{Automation}
Some artifacts benefit from continuous integration and continuous delivery (CI/CD) pipelines and automation, such as automated installation and experimentation. We believe such pipelines can improve authors' productivity and save time in the evaluation procedure~\cite{jimenez2017popperci}. We believe that this may address challenges \textbf{C2}, \textbf{C3}, \textbf{C4}, \textbf{C5}, \textbf{C6}, and \textbf{C7}.

\subsubsection{Problem and Experiment Dimensions}
Sometimes, artifact reviewers may need to read an accepted paper carefully to understand how to map the overall problem addressed by the paper to the experiments conducted using the artifact. Based on our experience, we suggest that charts explaining the experiments and problem dimensions should be included in both the paper and the artifact to help reviewers understand the different input parameters (experiment dimensions) under which the artifact is evaluated and how they map back to the problem the paper claims to solve.

\subsubsection{Public and Academic Clouds}
Some artifacts may require expensive or special resources that the reviewers cannot access. Consequently, some guidelines recommend authors provide access to such resources. This may be helpful if the evaluation is not double-blinded. However, it may be preferable for authors to use public or academic cloud resources and provide automation for such environments because AE committees will usually have some credits for public or academic clouds. We believe this may address challenges \textbf{C1} and \textbf{C7}.

\subsection{Recommendations for Artifact Evaluation Committees}

\subsubsection{Communication}
In our experience, artifact authors may inadvertently fail to describe some important elements in their documentation or paper, which may complicate the evaluation process. We have also found that a lack of communication between authors, reviewers, and chairs can slow down the progress of the evaluation. We therefore strongly recommend encouraging communication between all parties by having multiple stages of review (more than two) and increasing the size of the AE committee and the number of reviews per artifact. Ideally, widely used best practices for code review in software engineering (for example, workflows based on pull requests) would be applied because the current approach based on one or two rounds of review may be insufficient and inconsistent with these practices. 

\subsubsection{Bid Based on the Expertise}
Sometimes, a reviewer may be unfamiliar with the wider technology landscape surrounding an artifact, which may cause the evaluation to focus on the reviewer's ability to grasp the overall technology stack rather than the artifact itself. This may be particularly likely if the evaluation committee is small or when reviewers have limited experience with the subject of the work being evaluated. We, therefore, suggest that reviewers bid on an artifact only if they have experience with the relevant technologies and topic. Similarly, we have observed that in some AETs, the confidence level of the reviewers is not specified (in contrast to common practices for paper review), which could add further uncertainty when making final decisions about artifacts. We, therefore, suggest that the reviewer's confidence level in their evaluation should be included in the evaluation to resolve such issues.

\subsection{Recommendation for the Community}
Although it is clear that there is a strong desire to increase reproducibility within the distributed systems research community, there appears to be a lack of systematic ways of encouraging it, particularly for researchers in the early stages of their careers. We, therefore, suggest that all research policymakers encourage reproducibility and repeatability in various ways. For instance,  funding agencies could encourage and ask for artifacts to mitigate the current reproducibility crisis in open science. Scientific publishers could promote it by improving the reward mechanism and implementing evaluation platforms. The distributed research community could also play a key role by conducting research to support the development of systematic procedures for improving reproducibility, for example, by studying the benefits of crowd-sourcing techniques for AE. Finally, it should be acknowledged that currently established AE procedures vary widely and may cause confusion for researchers, especially as artifacts should ideally be constructed while the corresponding research is ongoing. Therefore, we strongly recommend forming a reproducibility working group to establish unified artifact guidelines for reviewers, authors, and chairs, such as the ACM Emerging Interest Group on reproducibility and replicability~\cite{eig}.

\section{Discussion}\label{discussion}

Reproducibility and repeatability are cornerstones of science. Despite their importance, various studies have shown a reproducibility crisis spanning multiple scientific fields and that many published studies cannot be replicated. The situation can be even worse in disturbed systems research, as work in this area may require access to large-scale and/or specialized hardware, and it is difficult to fully automate repeatable experiments in ways that make them accessible to anybody. While some strategies for promoting reproducibility have been developed within the distributed systems community, including artifact evaluation processes, our answer to the question of how unified existing artifact evaluation practices are in distributed systems research indicates that artifact evaluation practices in distributed systems research currently lag behind those used in other areas of computer science, are less unified, and are generally more complex (as shown in \cref{tab:criteria}). 

In this paper, we review the current guidelines of top conferences and journals in the distributed systems field and identify practices for artifacts and procedures for their evaluation for both artifact authors and evaluation committees to answer the question of what the current state of artifact evaluation practice is in the distributed systems field. 

To answer how reproducibility, artifacts, and artifact evaluation practices can be improved and promoted in distributed system conferences and journals, we propose additional recommendations for artifact evaluations to mitigate the challenges in distributed systems research. These suggestions are informed by our experience in best practices in software engineering. Furthermore, both the existing recommendations and those proposed in our study have the potential to be applied to research practices across diverse disciplines. Finally, we summarized all the existing and our proposed recommendations as a set of checklists for both artifact evaluation committees and artifact authors (see \cref{appendix-a,appendix-b}). In addition to our recommendations, we note that some communities have used crowd-sourcing for AE, and we believe that such approaches could also help to improve repeatability and reproducibility in distributed systems research. \cref{fig:chal_rec_mapping} illustrates the relationship between the challenges identified and the current state of practices, alongside our recommendations. Notably, some practices and recommendations serve as metadata, enhancing the long-term repeatability of research. Drawing inspiration from the social science research community's approach, where independent replication of results is employed to enhance the reliability of findings\cite{replication}, we propose a similar strategy to bolster reproducibility and repeatability in the domain of distributed systems research. We hope this paper will stimulate further discussion within the distributed systems community regarding enhancing research reproducibility through more robust artifact practices and streamlined evaluation procedures.

\begin{acks}
This work was partially supported by the Wallenberg AI, Autonomous Systems and Software Program (WASP) funded by the Knut and Alice Wallenberg Foundation, and by the Swedish Research Council (VR), with the project ``PSI''.
\end{acks}
\balance
\bibliographystyle{acm}
\bibliography{references}

\begin{thebibliography}{10}

\bibitem{abedi2015conducting}
{\sc Abedi, A., Heard, A., and Brecht, T.}
\newblock Conducting repeatable experiments and fair comparisons using 802.11 n mimo networks.
\newblock {\em ACM SIGOPS Operating Systems Review 49}, 1 (2015), 41--50.

\bibitem{acm_AE_policy}
{\sc ACM}.
\newblock Artifact review and badging version 1.1.
\newblock Online, Aug 2020.

\bibitem{acm_dl}
{\sc ACM}.
\newblock Acm digital library, 2022.

\bibitem{niso2021}
{\sc Badging, N.~R., and Group, D.~W.}
\newblock Reproducibility badging and definitions, 2022.

\bibitem{bajpai2019dagstuhl}
{\sc Bajpai, V., Brunstrom, A., Feldmann, A., Kellerer, W., Pras, A., Schulzrinne, H., Smaragdakis, G., Wählisch, M., and Wehrle, K.}
\newblock The dagstuhl beginners guide to reproducibility for experimental networking research, 2019.

\bibitem{Baker2016}
{\sc Baker, M.}
\newblock 1,500 scientists lift the lid on reproducibility.
\newblock {\em Nature 533}, 7604 (May 2016), 452--454.

\bibitem{baldassarre2023re}
{\sc Baldassarre, M.~T., Ernst, N., Hermann, B., Menzies, T., and Yedida, R.}
\newblock (re) use of research results (is rampant).
\newblock {\em Communications of the ACM 66}, 2 (2023), 75--81.

\bibitem{barowy2022howtoAEC}
{\sc Barowy, D., Curtsinger, C., Tosch, E., and Vilk, J.}
\newblock Howto for aec submitters, 2022.

\bibitem{berger2019checklist}
{\sc Berger, E.~D., Blackburn, S.~M., Hauswirth, M., and W., H.~M.}
\newblock A checklist manifesto for empirical evaluation: A preemptive strike against a replication crisis in computer science, 2022.

\bibitem{chen2022towards}
{\sc Chen, B., Wen, M., Shi, Y., Lin, D., Rajbahadur, G.~K., and Jiang, Z.~M.}
\newblock Towards training reproducible deep learning models.
\newblock In {\em 2022 IEEE/ACM 44th International Conference on Software Engineering (ICSE)\/} (USA, 2022), IEEE, IEEE, pp.~2202--2214.

\bibitem{childers2017artifact}
{\sc Childers, B.~R., and Chrysanthis, P.~K.}
\newblock Artifact evaluation: Is it a real incentive?
\newblock In {\em 2017 IEEE 13th international conference on e-science (e-Science)\/} (USA, 2017), IEEE, IEEE, pp.~488--489.

\bibitem{childers2016artifact}
{\sc Childers, B.~R., Fursin, G., Krishnamurthi, S., and Zeller, A.}
\newblock Artifact evaluation for publications (dagstuhl perspectives workshop 15452).
\newblock In {\em Dagstuhl Reports\/} (Dagstuhl, Germany, 2016), vol.~5, Schloss Dagstuhl-Leibniz-Zentrum fuer Informatik, Schloss Dagstuhl--Leibniz-Zentrum fuer Informatik, pp.~29--35.

\bibitem{open2015estimating}
{\sc Collaboration, O.~S.}
\newblock Estimating the reproducibility of psychological science.
\newblock {\em Science 349}, 6251 (2015), aac4716.

\bibitem{collberg2015repeatability}
{\sc Collberg, C., Proebsting, T., and Warren, A.~M.}
\newblock Repeatability and benefaction in computer systems research.
\newblock {\em University of Arizona TR 14}, 4 (2015), 1–68.

\bibitem{repeatability_in_cs}
{\sc Collberg, C., and Proebsting, T.~A.}
\newblock Repeatability in computer systems research.
\newblock {\em Commun. ACM 59}, 3 (feb 2016), 62–69.

\bibitem{collberg2016repeatability}
{\sc Collberg, C., and Proebsting, T.~A.}
\newblock Repeatability in computer systems research.
\newblock {\em Communications of the ACM 59}, 3 (2016), 62--69.

\bibitem{ctuning}
{\sc cTuning}.
\newblock Reproducible papers with artifacts.
\newblock \url{https://ctuning.org/ae/}, 2022.

\bibitem{fursin2022ctuning}
{\sc cTuning Foundation}.
\newblock Artifact checklist.
\newblock \url{https://ctuning.org/ae/checklist.html}, 2022.

\bibitem{delling_et_al:DR:2016:6146}
{\sc Delling, D., Demetrescu, C., S.~Johnson, D., and Vitek, J.}
\newblock Rethinking experimental methods in computing (dagstuhl seminar 16111).
\newblock {\em Dagstuhl Reports 6}, 3 (2016), 24--43.

\bibitem{dietrich2015dataref}
{\sc Dietrich, C., and Lohmann, D.}
\newblock The dataref versuchung: Saving time through better internal repeatability.
\newblock {\em ACM SIGOPS Operating Systems Review 49}, 1 (2015), 51--60.

\bibitem{drummond2009replicability}
{\sc Drummond, C.}
\newblock Replicability is not reproducibility: nor is it good science.
\newblock In {\em Proceedings of the Evaluation Methods for Machine Learning Workshop at the 26th ICML\/} (USA, 2009), vol.~1, National Research Council of Canada Montreal, Canada, ACM.

\bibitem{dryad}
{\sc Dryad}.
\newblock Dryad - publish and preserve your data, 2022.

\bibitem{easterbrook2014open}
{\sc Easterbrook, S.~M.}
\newblock Open code for open science?
\newblock {\em Nature Geoscience 7}, 11 (2014), 779--781.

\bibitem{eig}
{\sc EIG, A.}
\newblock Acm eig on reproducibility and replicability, 2022.

\bibitem{fehr2016best}
{\sc Fehr, J., Heiland, J., Himpe, C., and Saak, J.}
\newblock Best practices for replicability, reproducibility and reusability of computer-based experiments exemplified by model reduction software.
\newblock {\em AIMS Mathematics 1}, 3 (2016), 261--281.

\bibitem{feitelson2015repeatability}
{\sc Feitelson, D.~G.}
\newblock From repeatability to reproducibility and corroboration.
\newblock {\em ACM SIGOPS Operating Systems Review 49}, 1 (2015), 3--11.

\bibitem{figshare}
{\sc figshare}.
\newblock figshare - credit for all your research, 2022.

\bibitem{Survey2018flittner}
{\sc Flittner, M., Mahfoudi, M.~N., Saucez, D., W\"{a}hlisch, M., Iannone, L., Bajpai, V., and Afanasyev, A.}
\newblock A survey on artifacts from conext, icn, imc, and sigcomm conferences in 2017.
\newblock {\em SIGCOMM Comput. Commun. Rev. 48}, 1 (apr 2018), 75–80.

\bibitem{fomel2008guest}
{\sc Fomel, S., and Claerbout, J.~F.}
\newblock Guest editors' introduction: reproducible research.
\newblock {\em Computing in Science \& Engineering 11}, 1 (2008), 5--7.

\bibitem{fritzsch2021resume}
{\sc Fritzsch, J., Wyrich, M., Bogner, J., and Wagner, S.}
\newblock R{\'e}sum{\'e}-driven development: a definition and empirical characterization.
\newblock In {\em Proceedings of the 43rd International Conference on Software Engineering: Software Engineering in Society\/} (USA, 2021), IEEE, pp.~19--28.

\bibitem{fursin2020enabling}
{\sc Fursin, G.}
\newblock Enabling reproducible ml and systems research: the good, the bad, and the ugly, 2020.

\bibitem{fursin2021collective}
{\sc Fursin, G.}
\newblock Collective knowledge: organizing research projects as a database of reusable components and portable workflows with common interfaces.
\newblock {\em Philosophical Transactions of the Royal Society A 379}, 2197 (2021), 20200211.

\bibitem{hermann2022has}
{\sc Hermann, B.}
\newblock What has artifact evaluation ever done for us?
\newblock {\em IEEE Security \& Privacy 20}, 5 (2022), 96--99.

\bibitem{hermann2020community}
{\sc Hermann, B., Winter, S., and Siegmund, J.}
\newblock Community expectations for research artifacts and evaluation processes.
\newblock In {\em Proceedings of the 28th ACM joint meeting on european software engineering conference and symposium on the foundations of software engineering\/} (2020), pp.~469--480.

\bibitem{heumuller2020publish}
{\sc Heum{\"u}ller, R., Nielebock, S., Kr{\"u}ger, J., and Ortmeier, F.}
\newblock Publish or perish, but do not forget your software artifacts.
\newblock {\em Empirical Software Engineering 25}, 6 (2020), 4585--4616.

\bibitem{ieee_xplore}
{\sc IEEE}.
\newblock Ieee xplore, 2022.

\bibitem{tpds_reproducibility}
{\sc {IEEE TPDS Reproducibility Review Board}}.
\newblock Tpds reproducibility initiative.
\newblock \url{https://www.computer.org/csdl/journal/td/misc/104303?title=Reproducibility%20Initiative&periodical=IEEE%20Transactions%20on%20Parallel%20and%20Distributed%20Systems}, 2024.

\bibitem{jasny2011again}
{\sc Jasny, B.~R., Chin, G., Chong, L., and Vignieri, S.}
\newblock Again, and again, and again…, 2011.

\bibitem{jimenez2017popperci}
{\sc Jimenez, I., Arpaci-Dusseau, A., Arpaci-Dusseau, R., Lofstead, J., Maltzahn, C., Mohror, K., and Ricci, R.}
\newblock Popperci: Automated reproducibility validation.
\newblock In {\em 2017 IEEE Conference on Computer Communications Workshops (INFOCOM WKSHPS)\/} (USA, 2017), IEEE, IEEE, pp.~450--455.

\bibitem{krishnamurthi2013artifact}
{\sc Krishnamurthi, S.}
\newblock Artifact evaluation for software conferences.
\newblock {\em ACM SIGSOFT Software Engineering Notes 38}, 3 (2013), 7--10.

\bibitem{software_crisis}
{\sc Krishnamurthi, S., and Vitek, J.}
\newblock The real software crisis: Repeatability as a core value.
\newblock {\em Commun. ACM 58}, 3 (feb 2015), 34–36.

\bibitem{Artifact2021Malik}
{\sc Malik, T.}
\newblock Artifact description/artifact evaluation: A reproducibility bane or a boon.
\newblock In {\em Proceedings of the 4th International Workshop on Practical Reproducible Evaluation of Computer Systems\/} (New York, NY, USA, 2021), P-RECS '21, Association for Computing Machinery, p.~1.

\bibitem{reproducibilty}
{\sc McNutt, M.}
\newblock Reproducibility.
\newblock {\em Science 343}, 6168 (2014), 229--229.

\bibitem{tdd-reproducibility}
{\sc Mugridge, R.}
\newblock Test driven development and the scientific method.
\newblock In {\em Proceedings of the Agile Development Conference, 2003. ADC 2003\/} (USA, 2003), IEEE, pp.~47--52.

\bibitem{national2019reproducibility}
{\sc of~Sciences~Engineering, N.~A., Medicine, et~al.}
\newblock Reproducibility and replicability in science, 2019.

\bibitem{papadopoulos2019methodological}
{\sc Papadopoulos, A.~V., Versluis, L., Bauer, A., Herbst, N., Von~Kistowski, J., Ali-Eldin, A., Abad, C.~L., Amaral, J.~N., Tuma, P., and Iosup, A.}
\newblock Methodological principles for reproducible performance evaluation in cloud computing.
\newblock {\em IEEE Transactions on Software Engineering 47}, 8 (2021), 1528--1543.

\bibitem{reproducibility_IEEE_CS}
{\sc Parashar, M.}
\newblock The reproducibility initiative.
\newblock {\em Computer 52}, 11 (nov 2019).

\bibitem{replication}
{\sc ReplicationWiki}.
\newblock Replicationwiki, 2022.

\bibitem{researchr}
{\sc Researchr}.
\newblock Researchr: a platform for finding, collecting, sharing, and reviewing scientific publications, 2022.

\bibitem{Saucez2018thoughts}
{\sc Saucez, D., and Iannone, L.}
\newblock Thoughts and recommendations from the acm sigcomm 2017 reproducibility workshop.
\newblock {\em SIGCOMM Comput. Commun. Rev. 48}, 1 (apr 2018), 70–74.

\bibitem{stodden2013best}
{\sc Stodden, V., and Miguez, S.}
\newblock Best practices for computational science: Software infrastructure and environments for reproducible and extensible research.
\newblock {\em Journal of Open Research Software 1}, 2 (2013), 1--6.

\bibitem{Timperley2021artifact-sharing}
{\sc Timperley, C.~S., Herckis, L., Le~Goues, C., and Hilton, M.}
\newblock Understanding and improving artifact sharing in software engineering research.
\newblock {\em Empirical Software Engineering 26}, 4 (2021), 67.

\bibitem{ds-definition}
{\sc van Steen, M., and Tanenbaum, A.~S.}
\newblock A brief introduction to distributed systems.
\newblock {\em Computing 98}, 10 (2016), 967--1009.

\bibitem{vaux2012replicates}
{\sc Vaux, D.~L., Fidler, F., and Cumming, G.}
\newblock Replicates and repeats—what is the difference and is it significant? a brief discussion of statistics and experimental design.
\newblock {\em EMBO reports 13}, 4 (2012), 291--296.

\bibitem{vivtek2011repeatability}
{\sc Vitek, J., and Kalibera, T.}
\newblock Repeatability, reproducibility, and rigor in systems research.
\newblock In {\em Proceedings of the Ninth ACM International Conference on Embedded Software\/} (New York, NY, USA, 2011), EMSOFT '11, Association for Computing Machinery, p.~33–38.

\bibitem{wikicfp}
{\sc WikiCFP}.
\newblock Wikicfp : Call for papers of conferences, workshops and journals, 2022.

\bibitem{wilkinson2019fair}
{\sc Wilkinson, M.~D., et~al.}
\newblock The {FAIR} guiding principles for scientific data management and stewardship.
\newblock {\em Scientific Data 3}, 1 (2016), 160018.

\bibitem{zenodo}
{\sc Zenodo}.
\newblock Zenodo - research. shared., 2022.

\end{thebibliography}
\newpage
\appendix
\nobalance

\section{Checklist for Artifact Authors}
\label{appendix-a}
In this appendix, we have compiled a comprehensive list of both our own and existing recommendations for artifact authors.

\begin{todolist}
    \item \textbf{Plan for artifact submission in advance:} Prepare artifacts well in advance, following previous conference guidelines, as authors often receive short notice of acceptance and limited time to address comments while rushing to submit.
    \item \textbf{Introduction:} Provide an informative introduction for the artifact, outlining its purpose and relevance to the research paper.
    \begin{todolist}
        \item \textbf{Problem and experiment dimensions:} Include explanatory charts in both the paper and the artifact introduction for better comprehension of experiment-context mapping.
        \item \textbf{Algorithm:} Include the algorithm description in the introduction for clarity.
        \item \textbf{Required resources:} Specify hardware and software requirements, including versions, for experiments in the introduction.
        \item \textbf{Required time:} Incorporate time requirements for conducting experiments on artifacts, especially for complex scenarios spanning extended periods in the introduction.
        \item \textbf{Expected output and results:} Specify the expected results and output format for better artifact evaluation in the introduction.
        \item \textbf{Metrics:} Include descriptions of all metrics used to assess experiment results, covering both low-level resource measurements and higher-level functionality-related data, including capacities and activity rates.
    
    \end{todolist}

    \item \textbf{Use version control systems:} Ensure version control system for artifacts to prevent loss of features during evaluation.
    \item \textbf{Use test-driven development methods:} Employ test-driven development methods when preparing artifacts. As bug fixes might introduce new issues during the evaluation process and this method can prevent them. 
    \item \textbf{Use automation:} Incorporate continuous integration and continuous delivery (CI/CD) pipelines and automation (e.g., automated installation and experimentation) to enhance productivity and streamline the evaluation process.
    \item \textbf{Public and academic clouds:} Use public or academic cloud platforms with automation, as these are often available and can facilitate the artifact evaluation process.
    \item \textbf{Archival repositories:} Ensure research data and artifacts are shared following FAIR principles in archival repositories such as Zenodo, Figshare, etc. 
    \item \textbf{Licenses:} Include artifact license information if sharing publicly.
    \item \textbf{Programs:} Include information about additional programs or benchmarks and if they are included or need to be downloaded.
    \item \textbf{Inputs:} Include instructions about essential inputs (e.g., data sets, models, configuration files).
    \item \textbf{Run-time states:} Consider and report any run-time sensitivities, such as attribute values and system states.
    \item \textbf{Experiments:} Document experiment setup and execution and offer clear guidelines for result replication and reproduction. Include procedures like shell scripts, user manual steps, or automated workflows as necessary.
    \item \textbf{Sample data:} Provide collected data during the experiments in the artifact and provide clear instructions for repeating the analysis when necessary.
\end{todolist}

\section{Checklist for Artifact Evaluation Committees}
\label{appendix-b}
In this appendix, you will find an extensive compilation of recommendations tailored to artifact evaluation committees. These recommendations are designed to assist both artifact evaluators and artifact evaluation track chairs.

\begin{todolist}
    \item \textbf{Multi-stage reviewing:} Implement \textit{at least} a two-stage reviewing process for artifact evaluation, with the first stage focused on identifying missing elements and the second stage for a thorough evaluation.
    \item \textbf{Providing guidelines:} Provide guidelines for both artifact evaluators and artifact authors.
    \begin{todolist}
        \item Create clear guidelines for artifact authors, detailing critical components and best practices to ensure the accessibility and understandability of their artifacts.
        \item Develop specific guidelines that define the evaluation criteria to enhance the accuracy of artifact assessments.
        \item Consider the varied expertise of committee members, including early career researchers, when formulating guidelines to improve artifact evaluation processes.
    \end{todolist}
    \item \textbf{Communication:} Ensure anonymous and open communication for reviewers and authors through a suitable platform to facilitate multi-stage artifact assessments and provide guidance to inexperienced reviewers. Encourage interaction between reviewers, authors, and chairs.
    \item \textbf{Bidding process:} Encourage reviewers to assess artifacts only if they possess experience with the related technologies and topics to prevent evaluations from focusing on the reviewer's grasp of the technology stack rather than the artifact itself. Recommend specifying the reviewer's confidence level in their own evaluation, which is the same as common practices for paper review, to reduce uncertainty and improve decision-making in artifact evaluation tracks.
\end{todolist}

\end{document}